\documentclass{aa}
\usepackage{graphics}
\begin{document}
\title{Lithium 6104{\AA} in Population II stars}

\author{A. Ford
\inst{1,2}
\and R.D. Jeffries
\inst{1}
\and B. Smalley
\inst{1}
\and S.G. Ryan
\inst{2}
\and W. Aoki
\inst{3}
\and S. Kawanomoto
\inst{3}
\and \nobreak{D.J. James}
\inst{4,5,6}
\and J.R. Barnes
\inst {4}}
\offprints{A. Ford}
\mail{alison.ford@open.ac.uk}
\institute{Department of Physics, Keele University, Keele, Staffordshire, ST5 5BG. UK
\and Department of Physics and Astronomy, The Open University, Walton Hall, Milton Keynes, MK7 6AA, UK
\and National Astronomical Observatory of Japan, Mitaka, Tokyo, 181-8588 Japan 
\and School of Physics and Astronomy, University of St Andrews, North Haugh, St Andrews, Fife, KY16 9SS, UK 
\and Laboratoire d'Astrophysique, Observatoire de Grenoble, BP-53, F-38041 Grenoble, Cedex 9, France 
\and Observatoire de Gen\`{e}ve, Chemin des Maillettes, $\#51$, Sauverny, CH-1290, Switzerland \\
email: alison.ford@open.ac.uk; rdj@astro.keele.ac.uk; bs@astro.keele.ac.uk; s.g.ryan@open.ac.uk; aoki.wako@noa.ac.jp; kawanomo@optik.mtk.nao.ac.jp; djames@obs.ujf-grenoble.fr}
\date{Received; accepted}
\authorrunning{A. Ford et al.}
\abstract
{We have obtained \'echelle spectroscopy of 14 Population II objects selected 
from those previously observed by Bonifacio \& Molaro (1997). 
For one object, HD~140283, we obtained exquisite data with the High Dispersion 
Spectrograph on the Subaru Telescope, with $S/N$ exceeding 1000 per 
0.018~\AA\ pixel. 
Li abundances 
have been determined by spectral synthesis from both the 6708{\AA} 
resonance line and also from 6104{\AA} subordinate feature. Firm 
detections of the weak line have been made in seven objects, and upper 
limits are reported for the remainder. 
Our 6708{\AA} abundances agree with those reported by Bonifacio \& Molaro at 
the 99\%-confidence level. Abundances from the 6104{\AA} line hint at a 
higher Li abundance than that determined from the resonance feature, 
but this evidence is mixed; the weakness of the 6104{\AA} line and the large number of upper limits make it difficult to draw firm conclusions. NLTE-corrections increase (rather than eliminate) the size of the (potential) discrepancy, and binarity appears unlikely to affect any abundance difference. The effect of multi-dimensional atmospheres on the line abundances was also considered, although it appears that use of 3-D (LTE) models could again act to {\em increase} the discrepancy, if one is indeed present.
}
\maketitle

\section{Introduction}
\label{sec-intro}

Population II stars are older and more metal-poor than the Pop. I stars which make up most of the Galactic disk. They are predominantly found in the Galactic bulge, in the spherical halo around the Galaxy, and in globular clusters, although they also make up $\sim$0.1\% of the population of nearby stars. They formed early in the life of the Galaxy, and therefore are better indicators of early Li abundances than their younger Pop. I counterparts. This seemed to be supported by the discovery of a Li abundance plateau in these objects by Spite \& Spite (\cite{spite82}). The Spite plateau (Li abundance\,=\,2.09\,$^{+0.19}_{-0.13}$\,dex, Ryan et al. \cite{ryan00} -- where A(Li) = $\log_{10}\frac{N({\mathrm{Li}})}{N({\mathrm{H}})} + 12$) includes objects in the temperature range 5600\,K\,$\leq T_{\mathrm{eff}} \leq$\,6400\,K, and metallicity range [Fe/H]$< -$1.3. Since the plateau Pop. II stars are predicted to have undergone little Li depletion over the course of their lives ({$<$0.1\,dex according to Ryan et al. 1999 --hereafter \cite{rnb99}--, $\leq$0.2\, dex according to Pinsonneault et al. \cite{p99}, but possibly as high as 0.3\,dex according to Salaris \& Weiss \cite{salaris01}}) the plateau Li abundance has been suggested to be close to the primordial Li abundance, Li$_{\mathrm{p}}$.
Recent work by \cite{rnb99} found that the plateau is ultra-thin, but slopes with metallicity ($\frac{\mathrm{d(A(Li))}}{\mathrm{d[Fe/H]}}\,=\,+0.118\pm0.023$), suggesting that the Galactic Li abundance increased slowly with metallicity during the time when Pop. II stars were forming. Li abundances in Pop. II stars cooler than the plateau are depleted. It is also possible that the plateau abundance is the result of stellar Li depletion from a higher value (see e.g. Pinsonneault \cite{p97}, \cite{p99}), though why the resulting Li abundance should then be so uniform in these objects is unclear (Th\'eado \& Vauclair \cite{theado01}). Salaris \& Weiss (2001) have developed models which include atomic diffusion (untempered by mass loss) which they compare with the plateau Li-$T_{\rm eff}$ trend. They conclude that this mechanism leaves the stars' Li abundances depleted by $\sim$0.3~dex relative to their initial abundances. They comment that the neglect of both mass-loss and radiative-levitation processes in their models probably leads to this being an overestimation, but more work is required to quantify this. However, the Spite plateau is probably still the best indicator available (after consideration of, and correction for, effects such as diffusion) for the primordial Li abundance, and can still be used to set limits on cosmological parameters such as $\eta$ (the baryon-to-photon ratio) and $\Omega_{\mathrm{B}}$ (the universal baryon density). 

There is yet another problem with assuming that the plateau abundance is the same as Li$_{\mathrm{p}}$: the plateau abundance is derived from only one spectral line, the 6708{\AA} resonance feature. Recent work by Kurucz (\cite{kurucz95}), Carlsson et al. (\cite{carlsson94}), Stuik et al. (\cite{stuik97}), Uitenbroek (\cite{uitenbroek98}) and Asplund et al. (\cite{asplund99}) has suggested that this line might be unrepresentative of the actual Li abundance within the stars, due to omissions in our understanding and modelling of stellar atmospheres. Recent works on halo stars (Bonifacio \& Molaro 1998, hereafter \cite{bm98} -- and next subsection) and young cluster stars (Soderblom et al. \cite{s93a}; Russell, \cite{russell96}; Ford et al. \cite{ford02}) have used the weak (excitation potential\,=\,1.8\,eV) \ion{Li}{i} subordinate line at 6104{\AA} in addition to the 6708{\AA} line (hereafter 6708 and 6104 are taken to mean `the \ion{Li}{i} line at 6708{\AA}' and `the \ion{Li}{i} line at 6104{\AA}' respectively), to attempt to determine Li abundances. This paper builds on those previous studies.

\subsection{Multi-dimensional atmospheres: the story so far...}
\begin{itemize}
\item Kurucz (\cite{kurucz95}) noted that one-dimensional model atmospheres will only work well if they have the same temporally- and spatially-averaged behaviour as the species they are used to predict, which might not be the case for Li. Kurucz considered the idea that conventional abundance analysis is in error, and suggested that model atmospheres could underestimate the amount of ionized Li by an order of magnitude, which would lead to Li abundance measurements also being lower than the `true' value by around 1\,dex. Kurucz cautions us that ``{\em since very few lines have atomic data known accurately enough to constrain the model, a match does not necessarily mean that the model is correct.}''
\item Carlsson et al. (\cite{carlsson94}) studied the effect of non-local thermodynamic equilibrium (NLTE) on the formation of Li lines. They used a standard flux-constant plane-parallel atmosphere to model Li line formation, but included processes ignored by assumptions of local thermodynamic equilibrium (LTE). Inclusion of NLTE processes appears to have some effect on Li lines. NLTE corrections are largest for cool ($T_{\mathrm{eff}} \sim $ 4500\,K), metal-poor ([Fe/H] $\sim$ $-$2) objects, although the magnitude of the effect varies with both temperature and metallicity. Generally, the corrections have opposite signs for 6104 and 6708, increasing the abundance derived from the subordinate line by up to 0.06\,dex (at A(Li) = 2.2) while decreasing it in the case of the resonance feature (by around 0.004\,dex at A(Li) = 2.2 for an object with $T_{\mathrm{eff}}$ = 6000\,K, $\log\,g$ = 4.0, and [Fe/H] = $-$2.0).
\item Stuik et al. (\cite{stuik97}) continued the work of Carlsson et al., investigating the sensitivity of \ion{Li}{i} and \ion{K}{i} to activity in Pleiades stars. They stress that the variations in Li line strength in the Pleiades cannot be attributed to a spread in the stellar abundances until such time as the line-formation predictions and atmospheric models can be shown to be accurate. One test of this is the use of the potassium resonance line (which should not show any abundance spread as K is not expected to be depleted in these stars) as a proxy for Li. This line is formed in similar conditions and transitions to the Li resonance feature, and so should be affected similarly. Their results suggest that while Li and K are not sensitive to the direct effects of chromospheric activity, they can be strongly affected by temperature variations deeper in the stellar atmosphere, and by interactions between regions of different temperature stratification. Such effects are not included in the conventional one-dimensional models, and therefore might account for some of the effects seen. Pop. II stars should not be as active as young cluster objects, but there are likely to be similar (related or unrelated) omissions in our modelling of Pop. II stars.
Kiselman (\cite{kiselman97, kiselman98}) used a 3-D solar-granulation snapshot to investigate the effects of departures from LTE in line formation on 6708. He reported marked line-strength variations over the granulation pattern for NLTE and LTE. He also produced syntheses of the line under these conditions and compared them to solar observations, finding that calculations which included the detailed modelling of the line radiative transfer matched observations better than those which neglected it.
\item Uitenbroek (\cite{uitenbroek98}) worked in a similar vein to Stuik et al. and Kiselman, considering the effect of convective surface inhomogeneities on the formation of the Li resonance line in the solar case. Uitenbroek uses 1.5- and two-dimensional NLTE calculations to investigate the effects of granulation on the lines. Again, granulation is unlikely to play a significant role in Pop. II-star Li modelling, but the fact that its omission in Pop. I models leads to a difference in abundance (although typically less that 0.1\,dex) shows that our understanding of stellar atmospheres is not complete.
\item BM98 detected 6104 in the Pop. II star \object{HD\,140283}. This feature is important since it is formed deeper in the atmosphere than the resonance line, and can therefore be used to test our models, since both lines should present the same abundance. Bonifacio \& Molaro's analysis suggested that the same abundance could be used to adequately reproduce both lines using only 1-D, homogeneous models. They took this to imply that these `simple' models were essentially correct. 
\item Asplund et al. (\cite{asplund99}) have used 3-D, time-dependent surface-convection simulations of two Pop. II objects, HD\,140283 and \object{HD\,84937} to investigate the effect of multi-dimensional modelling on Li abundance in metal-poor stars. They report that, as might be expected, three-dimensional model atmospheres have a different temperature structure to one-dimensional models. The effect of their 3-D, LTE models on the stellar Li is to decrease the abundance measured from the resonance line by 0.2-0.35\,dex, relative to that obtained using
1-D models (for models with $T_{\mathrm{eff}}$\,=\,5690\,K, 
$\log\,g$\,=\,3.67, [Fe/H]\,=\,$-$2.5 and microturbulence 
($\xi$)\,=\,1\,km\,s$^{-1}$ in the case of HD\,140283, and 
$T_{\mathrm{eff}}$\,=\,6330\,K, $\log\,g$\,=\,4.04, [Fe/H]\,=\,$-$2.25 and
$\xi$\,=\,1\,km\,s$^{-1}$ for HD\,84937). Note that Asplund et al. \cite{asplund99} quote $\log\,g$ for $g$ in units of m\,s$^{-1}$; for consistency with other works we use cm\,s$^{-1}$. 
However, 3-D NLTE corrections almost completely cancel the 3-D LTE effect, leading to $<$\,0.1\,dex change (in these two cases) to the 1-D LTE abundance (Asplund et al. \cite{asplund00}).
\end{itemize}
We seek here to extend the work of \cite{bm98}. Firstly, their sample consists of only one star, which might or might not be typical of other Pop. II objects. Secondly, they do not fit 6104 independently of the resonance feature, but synthesize it at the abundance determined from the stronger feature. 6104 is difficult to measure due to its weakness and proximity to Fe and Ca lines. At the signal-to-noise ratio ($S/N$) of the \cite{bm98} spectrum it is likely that the 6104 Li abundance can only be determined to $\pm$0.2\,dex (1$\sigma$). As this is comparable with the likely size of problems with the models and larger than the uncertainties claimed by some authors in Li$_{\mathrm{p}}$, a more sensitive comparison of 6708 and 6104 is desirable. Because of the weakness of the line, and difficulties achieving higher $S/N$, it is sensible to measure 6104 in a number of objects to reduce the statistical errors (see Sect.~\ref{sec-abundanceerrors} and~\ref{sec-EWerrors}).

Following on from the work of \cite{bm98} we have obtained a larger sample of plateau stars (and three cooler objects) and analysed them with a variety of 1-D homogeneous models, fitting both lines independently of each other to determine if the abundances from the two lines do agree.

\section{Observations and Reduction}
\label{sec-obsred}

Our sample consisted of 14 Pop. II stars taken from the larger sample of Bonifacio \& Molaro (1997 -- hereafter \cite{bm97}) with [Fe/H] from $-$1 to $-$2.8, and temperatures in the range 5400\,K $\leq T_{\mathrm{eff}} \leq$ 6330\,K (as quoted by \cite{bm97}). These are listed in Table~\ref{table-targets}.

\begin{table*}
\caption{Identifiers, effective temperatures, surface gravities and metallicities for sample stars. Variations in $S/N$ between the orders arise from different numbers of counts in each \'echelle spectral order and the presence of cosmic rays near the lines in some objects, resulting in fewer exposures being co-added for some objects. Note that the identifiers in bold are those used throughout this paper.}
\begin{tabular}{llllllllllll}
\hline
Star Identifier & & & Other & $T_{\mathrm{eff}}$ & $\log\,g$ & [Fe/H] & S/N & S/N & $v \sin i^{\star}$ & Run$^{\dag}$ & Notes \\
 & & & & & & & 6708 & 6104 &(km\,s$^{-1}$) & & \\ 
\hline
{\bf \object{HD\,84937}} & \object{G43-003} & \object{BD+14 2151} &                      & 6330 & 3.90 & $-$2.49 & 285  & 300 & 7   & 6 & CS \\
\object{HD\,83769C} & \object{G49-029} & \object{BD+01 2341} & {\bf \object{LP 608-62}}  & 6313 & 3.90 & $-$2.81 & 240  & 220 & 5   & 6 & Cs \\
{\bf \object{HD\,74000}} &  & \object{BD-15 2546} &                                      & 6224 & 4.08 & $-$2.00 & 230  & 249 & 7   & 5 &    \\
{\bf \object{HD\,108177}} & \object{G13-035} & \object{BD+02 2538} &                     & 6097 & 4.03 & $-$1.98 & 250  & 250 & 5   & 6 & s  \\
\object{HD\,28428} & \object{G8-016} & {\bf \object{BD+21 607}} &                        & 6034 & 4.05 & $-$1.61 & 220  & 430 & 6   & 6 & s  \\
{\bf \object{HD\,194598}} & \object{G24-015} & \object{BD+09 4529} &                     & 6017 & 4.15 & $-$1.37 & 160  & 240 & 4.5 & 2 &    \\
{\bf \object{HD\,219617}} & \object{G273-011} & \object{BD-14 6437} &                    & 6012 & 4.30 & $-$1.63 & 260  & 240 & 6.5 & 2,3,4 & C  \\
{\bf \object{HD\,94028}} & \object{G55-025} & \object{BD+21 2247} &                      & 6001 & 4.15 & $-$1.50 & 270  & 270 & 4.5 & 3,5 & S  \\
{\bf \object{HD 201891}} & & \object{BD+17 4519} &                                       & 5909 & 4.18 & $-$1.22 & 285  & 250 & 4   & 2,3 & S  \\
\object{HD\,24289} & {\bf \object{G80-28}} & \object{BD-04 680} &                        & 5866 & 3.73 & $-$2.07 & 250  & 330 & 5   & 6 &    \\
{\bf \object{HD\,193901}} &  & \object{BD-21 5703} &                                     & 5750 & 4.40 & $-$1.13 & 130  & 180 & 3.5 & 2,4 &    \\
{\bf \object{HD\,140283}} &   & \object{BD-10 4149} &                                     & 5691 & 3.35 & $-$2.37 & 1100 & 1100& 4  & 7 &    \\ 
{\bf \object{HD\,188510}} & \object{G142-017} & \object{BD-10 4091} &                    & 5564 & 4.88 & $-$1.80 & 250  & 255 & 2   & 2,3 & CS \\
{\bf{HD\,64090}} & \object{G90-025} & \object{BD+31 1684} &                              & 5441 & 4.50 & $-$1.82 & 300  & 370 & 1   & 3,5 &    \\
\hline
\multicolumn{12}{l}{$\star$ $v \sin i$ has been used to model macroturbulence, and does not imply that the stars rotate at these speeds.}\\
\multicolumn{12}{l}{$\dag$ The relevant Table~\ref{table-obs} column numbers for each object.}\\
\multicolumn{12}{l}{Notes: C -- classified as a binary or suspected binary by Carney (\cite{carney83})} \\
\multicolumn{12}{l}{S -- classified as a definite binary by Stryker et al. (\cite{stryker85})} \\
\multicolumn{12}{l}{s -- classified as near significance criterion for binaries by Stryker et al. (\cite{stryker85}).}\\
\end{tabular}
\label{table-targets}
\end{table*}

Spectra were taken on the 4.2m William Herschel Telescope (WHT) on 1998 October
08 and 22, 1998 November 29, 1999 December 22-23, 2001 January 23-26 and 2001 
April 26, on the 4m Anglo-Australian Telescope (AAT) on 1999 October 22, and 2001 
April 26, and on the 8.2m Subaru Telescope on 2001 July 22. 
The Utrecht \'Echelle Spectrograph (on the WHT) and UCL \'Echelle Spectrograph (on the AAT) were used, both with E31 gratings, for their high resolving power ($R\sim$50000). 
On the Subaru Telescope, the High Dispersion Spectrograph was used, configured 
for $R\sim90000$. 
The usual calibration frames were taken in all cases, including tungsten flat 
fields, bias frames, and thorium-argon hollow-cathode-lamp spectra. The data 
were extracted and wavelength calibrated  using the Starlink {\sc echomop} (Mills et al. \cite{mills97}) and 
{\sc figaro} (Shortridge et al. \cite{shortridge99}) packages for the WHT and AAT data,
and IRAF for the Subaru data. Scattered-light subtraction was included in the data reduction. 
Several exposures of each target were taken and co-added to give a typical $S/N$ of 250-350 per pixel, but very much higher quality data were obtained for HD~140283 (13 observations were taken, with a total exposure time of 4924 seconds), where we achieved $S/N$~=~1100 per 0.018~\AA\ pixel! The spectrum was heavily affected by fringing for wavelengths $>$\,7000{\AA}, but this was not a problem in the region of the lithium lines.

\begin{figure}
\resizebox{7.5cm}{!}{\includegraphics{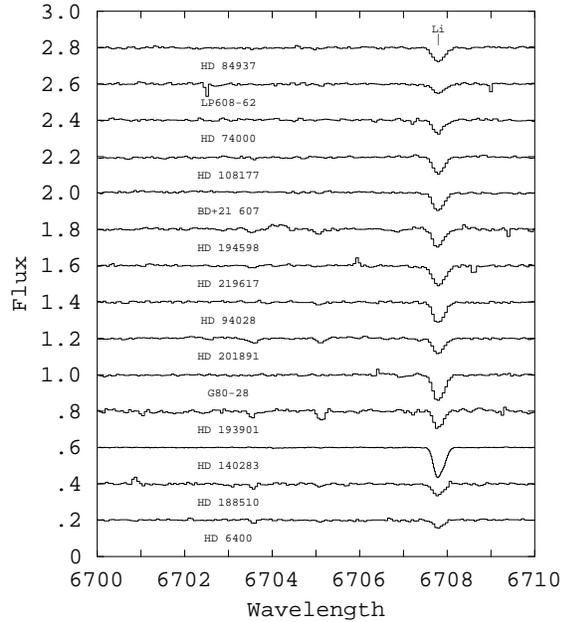}}
\caption{Normalised spectra for our sample stars in the region of the 6708 line. The spectra have been vertically offset for clarity. \ion{Fe}{i} lines at 6703{\AA} and 6705{\AA} can be seen in the more metal-rich objects.}
\label{fig-spec7}
\end{figure}

\begin{figure}
\resizebox{!}{21cm}{\includegraphics{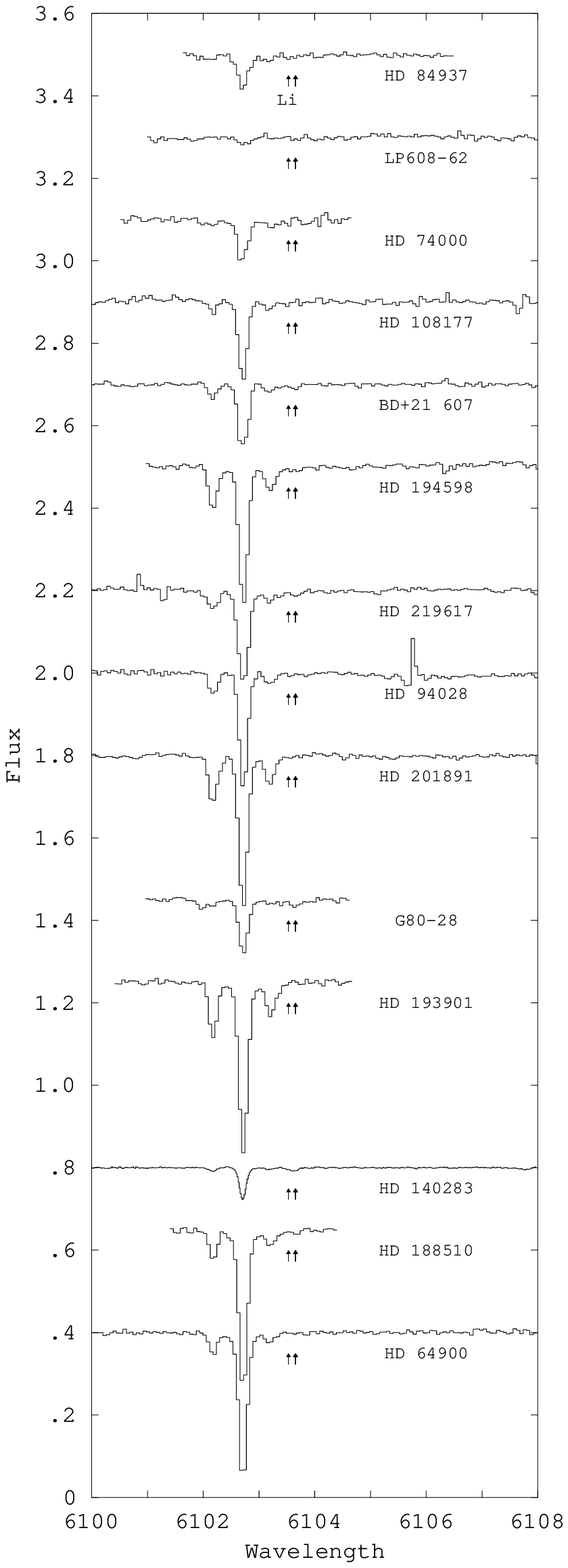}}
\caption{Normalised spectra for our sample stars in the region of 6104. Arrows indicate the wavelengths of the 6104 triplet, which can be seen in {\bf{HD\,84937}}, BD+21 607, HD\,194598, HD\,219617, G80-28, HD\,140283 and HD\,188510. The spectra have been vertically offset for clarity.}
\label{fig-spec4}
\end{figure}

These $S/N$ values were determined from the $RMS$ discrepancies arising from a 
polynomial fit to line-free spectral regions (as determined by a spectrum 
synthesis with appropriate $T_{\mathrm{eff}}$ and [Fe/H] parameters). This 
estimate of $S/N$ is likely to be representative, and up to $\sim$30\% smaller 
than $S/N$ determined from the total number of counts for each object. 
Fig.~\ref{fig-spec7} shows spectra in the region of 6708, while the 6104 spectra
are presented in Fig.~\ref{fig-spec4}. 

\begin{table*}
\caption{Observational information for data in this paper.}
\begin{tabular}{lllllll}
\hline
 (1) & (2) & (3) & (4) & (5) & (6) & (7) \\
Date & 1998 Oct & 1998 Nov & 1999 Oct & 1999 Dec & 2000 Jan & 2001 Jul\\
\hline
Telescope                 & WHT   & WHT   & AAT    & WHT   & WHT    & Subaru  \\
Detector                  & SITe1 & SITe1 & MITLL2 & SITe1 & SITe1  & 2$\times$EEV\\
Pixel size ($\mu$m)       & 22.5  & 22.5  & 15     & 22.5  & 22.5   & 13.5    \\
Dispersion ({\AA}/pixel)  & 0.05  & 0.05  & 0.05   & 0.06  & 0.05   & 0.02    \\
Slit width                & 1.2'' & 1.3'' & 1.2''  & 1.4'' & 1.16'' & 0.4''   \\
FWHM at 6104{\AA} ({\AA}) & 0.12  & 0.12  & 0.12   & 0.16  & 0.13   & 0.07    \\
\hline
\end{tabular}
\label{table-obs}
\end{table*}

\section{Analysis}
\label{sec-analysis}

$T_{\mathrm{eff}}$, $\log\,g$, and [Fe/H] values were taken from \cite{bm97}. These values are presented along with alternate identifiers in Table~\ref{table-targets}. A microturbulence of 1\,km\,s$^{-1}$ was used, as in \cite{bm97} and \cite{bm98}. The analysis is virtually independent of the microturbulence value used; a change of +0.5\,km\,s$^{-1}$ affects 6708 by less than 0.04 dex, and 6104 by less than 0.01 dex, for all objects in our sample). Initially a $v \sin i$ parameter of 5\,km\,s$^{-1}$ was used for all the objects, but this was varied until the best $\chi^2$ value for the fit was obtained. We have used the $v \sin i$ parameter to model macroturbulent broadening of the lines, rather than as a physical rotational velocity. Instrumental broadening, determined by fitting Gaussians to arc lines, was also included in the syntheses. The values of both $v \sin i$ and instrumental broadening parameters used are included in Table~\ref{table-targets}. The models used were Kurucz 1-D, homogeneous, LTE, {\sc atlas9} model atmospheres with the mixing-length theory of convection (MLT -- Castelli et al. \cite{castelli97}) with overshooting turned off.
 
Li abundances and EWs were determined by fitting a synthetic spectrum to the data using {\sc uclsyn} (Smith \cite{smith92}, Smalley et al. \cite{smalley01}) and the line list in Table~\ref{table-linelist}. The synthesis for 6708 included the weak \ion{Fe}{i} line in the blue wing of the resonance feature. Given the low metallicities of these stars, the Fe line is unlikely to have a significant EW, but it was included for completeness. The abundances of Fe and Ca were determined from the lines near 6104 by fitting syntheses to them prior to fitting the Li line. The $gf$ factors of these additional lines were fixed by matching the lines in our observed and synthesized solar spectra. Best fits were determined by $\chi^2$ minimization. The solar model used was: $T_{\mathrm{eff}}$\,=\,5777\,K, $\log\,g$\,=\,4.44, $\xi$\,=\,1.5\,km\,s$^{-1}$. The Li $gf$ factors are those contained in the Vienna Atomic Line Database (VALD - Piskunov et al. 
\cite{piskunov95}). The VALD data for 6104 is also identical to that presented by Lindgard \& Nielsen (\cite{lindgard77}), which are accurate to within 10\% for 6104, and within 3\% for 6708. The Van der Waals broadening coefficients for the \ion{Fe}{i} and \ion{Fe}{ii}, \ion{Ca}{i} and \ion{Li}{i} lines were those reported by Barklem et al. (\cite{barklem00} -- hereafter ABO values).

\begin{table}[h]
\caption{Lines used in spectral synthesis and analysis.}
\begin{tabular}{lllr}
\hline
Element & Wavelength & Excitation & $\log\,gf$ \\
        & ({\AA})    & Potential (eV) &       \\
\hline
Fe I & 6707.432 & 4.608 & $-$2.357  \\
Li I & 6707.754 & 0.000 & $-$0.431  \\
Li I & 6707.766 & 0.000 & $-$0.209  \\
Li I & 6707.904 & 0.000 & $-$0.773  \\
Li I & 6707.917 & 0.000 & $-$0.510  \\
\hline
Fe I & 6102.159 & 4.608 & $-$2.692 \\  
Fe I & 6102.173 & 4.835 & $-$0.454  \\
Ca I & 6102.723 & 1.879 & $-$0.950  \\
Fe I & 6103.186 & 4.835 & $-$0.721  \\
Fe I & 6103.294 & 4.733 & $-$1.325  \\
Fe II& 6103.496 & 6.217 & $-$2.224  \\
Li I & 6103.538 & 1.848 & 0.101  \\
Li I & 6103.649 & 1.848 & 0.361  \\
Li I & 6103.664 & 1.848 & $-$0.599 \\
\hline
\end{tabular}
\label{table-linelist}
\end{table}

6104 can be seen in the spectra of some of our sample (Fig.~\ref{fig-spec4}): HD\,84937, BD+21 607, HD\,194598, HD\,219617, G80-28, HD\,140283 and HD\,188510 all show a feature at the appropriate wavelengths. For the other objects a 3$\sigma$ upper limit was determined from the pixel size ($p$, in {\AA}), $S/N$, and the physical width of the line ($r$, in {\AA}) in other objects with similar atmospheric parameters such that: $\sigma_{EW}$ = $\frac{\surd{(rp)}}{S/N}$. The corresponding abundance limit was determined by performing a synthesis with $EW$ = 3$\sigma_{EW}$.

\begin{figure}
\resizebox{!}{!}{\includegraphics{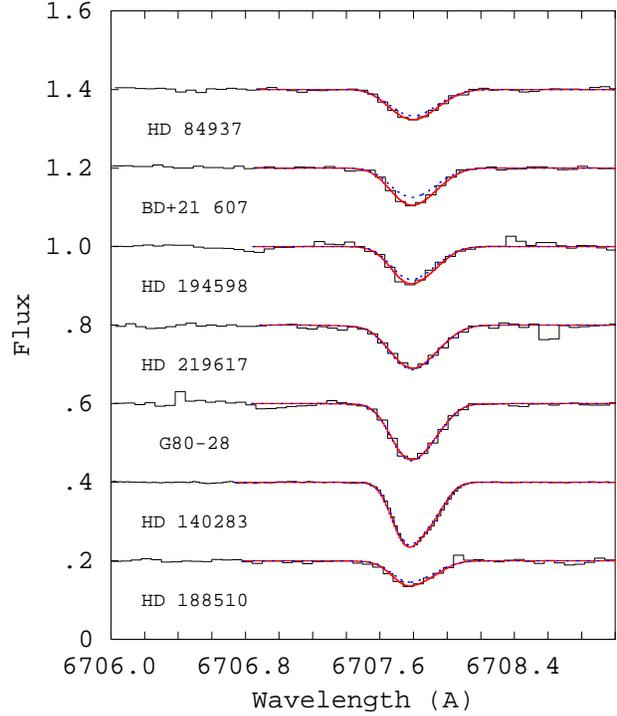}}
\caption{Observed data (solid lines) and synthetic fits in the region of 6708, using our abundances (heavy lines) and the abundances reported in \cite{bm97} (dotted lines) for objects with 6104 detections. Spectra are vertically offset for clarity.}
\label{fig-specfit7}
\end{figure}

\begin{figure}
\resizebox{!}{21cm}{\includegraphics{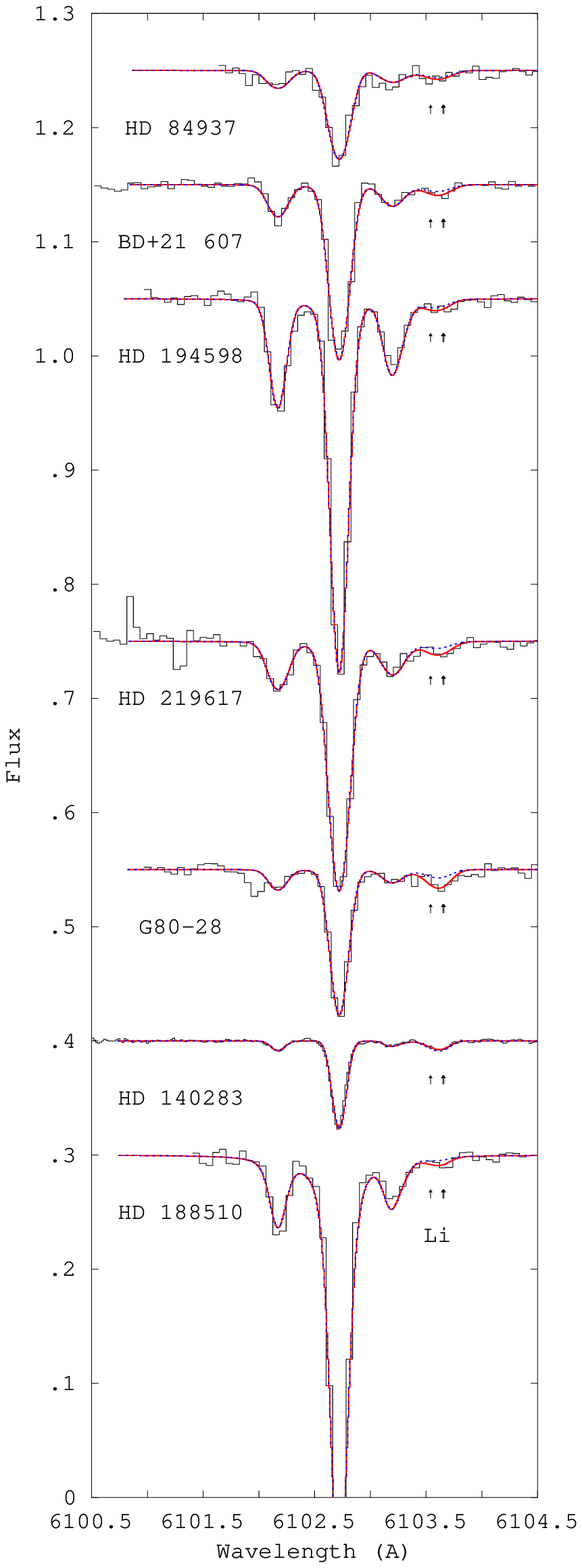}}
\caption{Observed data (histogram) and synthetic fits in the region of 6104, using our abundances (heavy lines) for objects with 6104 detections. Syntheses using the 6708 abundance are also shown (dashed lines). Spectra are vertically offset for clarity, and arrows indicate the position of the Li-triplet components (the arrows for 6103.65 and 6103.66 are superimposed).}
\label{fig-specfit4}
\end{figure}

Figs.~\ref{fig-specfit7} and ~\ref{fig-specfit4} show the detailed synthetic fits to the data in the 6708 and 6104 regions. Fig.~\ref{fig-specfit7} also shows the syntheses using our models and the abundances reported in \cite{bm97},
which we discuss later in the paper. 


\section{Results}
\label{sec-results}

LTE Li abundance measurements or upper limits for all the objects in our sample are presented in Table~\ref{table-results}, along with the synthetic EWs. We also corrected the LTE abundances for NLTE effects using the code of Carlsson et al. (\cite{carlsson94}), and these values are also presented in the table. At first glance the detections from 6104 appear higher than those from 6708, although there are upper limits where the abundances might agree. Before we can fully consider this as intrinsic to the stars, we need to consider possible sources of error which could account for the difference.

\begin{table*}
\caption{Li EWs and abundances for sample objects derived from 6708 and 6104.}
\begin{tabular}{llllrrr}
\hline
Object & EW$_{6708} \pm \sigma$ & A(Li)$_{6708} \pm \sigma$ & A(Li)$_{6708}$ & EW$_{6104} \pm \sigma$ & A(Li)$_{6104} \pm \sigma$ & A(Li)$_{6104}$\\
       & (m{\AA})    & LTE            & NLTE           & (m{\AA})    & LTE            & NLTE         \\
\hline
HD\,84937  & 26.67$\pm$0.78 & 2.36$\pm$0.01 & 2.30 & 2.24$\pm$0.77 & 2.46$^{+0.13}_{-0.18}$ & 2.50 \\
LP 608-62  & 16.71$\pm$1.01 & 2.09$\pm$0.03 & 2.07 & $\leq$1.98    & $\leq$2.41             & $\leq$2.46 \\
HD\,74000  & 22.22$\pm$0.90 & 2.18$\pm$0.02 & 2.18 & $\leq$1.93    & $\leq$2.37             & $\leq$2.42 \\
HD\,108177 & 29.81$\pm$0.98 & 2.23$\pm$0.02 & 2.22 & $\leq$1.80    & $\leq$2.28             & $\leq$2.32 \\
BD+21 607  & 31.55$\pm$0.83 & 2.23$\pm$0.02 & 2.23 & 2.62$\pm$0.68 & 2.44$^{+0.10}_{-0.13}$ & 2.50 \\
HD\,194598 & 30.28$\pm$0.81 & 2.21$\pm$0.02 & 2.21 & 2.55$\pm$0.94 & 2.43$^{+0.14}_{-0.20}$ & 2.49 \\
HD\,219617 & 38.28$\pm$0.60 & 2.31$\pm$0.02 & 2.30 & 3.36$\pm$0.67 & 2.54$^{+0.08}_{-0.10}$ & 2.60 \\
HD\,94028  & 36.51$\pm$0.88 & 2.29$\pm$0.02 & 2.29 & $\leq$1.61    & $\leq$2.21             & $\leq$2.27 \\
HD\,201891 & 26.31$\pm$0.83 & 2.07$\pm$0.02 & 2.08 & $\leq$1.74    & $\leq$2.21             & $\leq$2.28 \\
G80-28     & 46.21$\pm$0.52 & 2.28$\pm$0.01 & 2.29 & 4.41$\pm$0.60 & 2.57$^{+0.06}_{-0.06}$ & 2.65 \\
HD\,193901 & 26.48$\pm$1.75 & 1.96$\pm$0.03 & 1.99 & $\leq$3.45    & $\leq$2.47             & $\leq$2.53 \\
HD\'140283 & 48.10$\pm$0.11 & 2.157$\pm$0.002& 2.19 & 1.81$\pm$0.15 & 2.09$^{+0.04}_{-0.04}$ & 2.19 \\ 
HD\,188510 & 20.41$\pm$0.56 & 1.69$\pm$0.02 &$\star$1.74&2.08$\pm$0.63&2.17$^{+0.12}_{-0.16}$&$\star$2.23 \\
HD\,64090  & 12.05$\pm$0.92 & 1.32$\pm$0.03 & 1.38 & $\leq$1.68    & $\leq$1.96             & $\leq$2.06 \\
\hline
\multicolumn{7}{l}{$\star$ The $\log\,g$ values for HD\,188510 are out of range for the NLTE-correction script, so a $\log\,g$} \\
\multicolumn{7}{l}{of 4.5 was used. This is likely to make a difference of $\leq$0.01\,dex in abundance from either line.}\\ 
\end{tabular}
\label{table-results}
\end{table*}

\subsection{Atmospheric Uncertainties}
\label{sec-abundanceerrors}
Deblending is not likely to contribute significantly to the $EW$ uncertainties, except perhaps in the most metal-rich objects in our sample. The main factor in determining the quoted errors is the $S/N$ value adopted for each observation, in that those objects with low $S/N$ have larger errors or $EW$ upper-limit predictions. There appears to be a small spread in abundances from both lines (covering a range of 0.4\,dex for 6708 and $\geq$\,0.4\,dex for 6104, neglecting stars cooler than 5800\,K). We caution the reader that the uncertainties reported and plotted
are purely those intrinsic to the $S/N$ and the modelling/synthesis processes, 
determined by requiring the reduced-$\chi^{2}$ value to increase by 1, or the values from $\frac{\surd{(rp)}}{S/N}$ if greater (only the latter estimate is used where there is no detection of the line).
They do not include contributions for uncertainties in photometry, 
$T_{\mathrm{eff}}$, $\log\,g$, [Fe/H], or other parameters. Of these only the temperature error is likely to be significant, with a 100\,K increase in temperature raising the abundance by 0.07\,dex for 6708, or 0.04\,dex for 6104 (for a 6000\,K star with $\log\,g$\,=\,4.0, [Fe/H]\,=\,$-$2.0 and A(Li)\,=\,2.17). This would introduce an additional uncertainty of $\pm$0.03\,dex into the relative abundance from the two lines at 6000\,K, growing to 0.05\,dex at 5000\,K. A 0.5\,dex change in $\log\,g$ will raise the abundance by $<$0.01\,dex for both lines; increasing [Fe/H] by 0.5\,dex will only increase the abundance by $\sim$0.02\,dex. Neither of these can induce a large systematic discrepancy in the abundances measured from the two lines. \cite{bm97} report that a 0.5\,km\,s$^{-1}$ change in microturbulence will alter the abundance by only 0.005\,dex. Since the main aims of this work are to compare the abundances from the two lines, and to compare our results to those of \cite{bm97} and \cite{bm98} using {\it their} model parameters, the errors we have obtained should be sufficient. If our data are to be compared to those of other authors, the additional uncertainties should be taken into account. 

\subsection{Atomic Parameter Uncertainties}
\label{sec-EWerrors}
There will also be uncertainties due to the atomic parameters used, e.g. oscillator strengths. The errors in the $gf$ values we have used propagate to abundance differences of $\sim$0.01\,dex for 6708 and $\sim$0.04\,dex for 6104, leading to a differential uncertainty of $\sim$0.04\,dex. For the purpose of comparing our data with that of \cite{bm97}, these are unlikely to be important since we have used essentially the same line lists. Broadening parameters (e.g. Stark and collisional broadening factors) will probably vary between this analysis and others since we have used the ABO collisional broadening factors (Barklem et al. \cite{barklem00}) for the Fe, Ca and Li lines. The Stark factors are less important for lines which form in the outer layers of the atmosphere, as the Li lines do. The effect of using the ABO values rather than others is of order $-$0.01\,dex for objects with $T_{\mathrm{eff}}$\,=\,5000-6000\,K and [Fe/H]\,=\,$-$2.0. We are content that for the purposes of comparing our data with that of Bonifacio \& Molaro, and indeed other authors, our atomic data are satisfactory.

\subsection{Comparison with data presented by Bonifacio \& Molaro (\cite{bm97})}
\label{sec-bmcomparison}

Table~\ref{table-bmcompare} contains 6708 results from both this paper and \cite{bm97}.

\begin{table*}
\caption{Synthetic EWs and LTE-derived abundances for 6708 from this paper and \cite{bm97}.}
\begin{tabular}{lllllll}
\hline
 & \multicolumn{2}{c}{This paper} & \multicolumn{2}{c}{\cite{bm97}} \\
Object & EW & A(Li) & EW & A(Li) \\
\hline
HD\,84937  & 26.67$\pm$0.78 & 2.36$\pm$0.01 & 24.0$\pm$1.2 & 2.26$\pm$0.03 \\ 
LP 608-62  & 16.71$\pm$1.01 & 2.09$\pm$0.03 & 22.0$\pm$2.3 & 2.20$\pm$0.05 \\
HD\,74000  & 22.22$\pm$0.90 & 2.18$\pm$0.02 & 25.0$\pm$3.4 & 2.22$\pm$0.02 \\
HD\,108177 & 29.81$\pm$0.98 & 2.23$\pm$0.02 & 32.0$\pm$1.3 & 2.26$\pm$0.02 \\
BD+21 607  & 31.55$\pm$0.83 & 2.23$\pm$0.02 & 25.0$\pm$2.2 & 2.11$\pm$0.05 \\
HD\,194598 & 30.28$\pm$0.81 & 2.21$\pm$0.02 & 28.0$\pm$0.9 & 2.15$\pm$0.02 \\
HD\,219617 & 38.28$\pm$0.60 & 2.31$\pm$0.02 & 40.0$\pm$1.3 & 2.33$\pm$0.02 \\
HD\,94028  & 36.51$\pm$0.88 & 2.29$\pm$0.02 & 35.0$\pm$1.3 & 2.25$\pm$0.02 \\
HD\,201891 & 26.31$\pm$0.83 & 2.07$\pm$0.02 & 24.0$\pm$0.8 & 2.00$\pm$0.02 \\
G80-28     & 46.21$\pm$0.52 & 2.28$\pm$0.01 & 47.0$\pm$2.1 & 2.29$\pm$0.03 \\
HD\,193901 & 26.48$\pm$1.75 & 1.96$\pm$0.03 & 30.0$\pm$3.4 & 1.98$\pm$0.06 \\
HD\,140283 & 48.10$\pm$0.11 & 2.16$\pm$0.01 & 47.5$\pm$0.6 & 2.14$\pm$0.05 \\ 
HD\,188510 & 20.41$\pm$0.56 & 1.69$\pm$0.02 & 18.0$\pm$3.4 & 1.60$\pm$0.10 \\
HD\,64090  & 12.05$\pm$0.92 & 1.32$\pm$0.03 & 12.0$\pm$1.0 & 1.30$\pm$0.04 \\
\hline
\end{tabular}
\label{table-bmcompare}
\end{table*}

There is no systematic trend between either our EWs or abundances and those determined by \cite{bm97}. All our results agree within 3$\sigma$, and most within 2$\sigma$. This is not surprising; the main difference between the two analyses, besides the different spectra, is \cite{bm97}'s use of $\alpha$-enhanced opacities (with the $\alpha$-process elements -- O, Ne, Mg, Si, S, Ar, Ca and Ti -- enhanced by 0.4\,dex) which they suggest are more realistic for Pop.\,II stars; we used solar-scaled (non-enhanced) opacities as these were more readily available. We have determined that the difference in opacities affects the abundances of both lines by less than 0.03\,dex at 5000\,K (for a model with $\log\,g$\,=\,4.0, [Fe/H]\,=\,$-$1, $\xi$\,=\,2\,km\,s$^{-1}$, and EW$_{6104}$\,=\,5\,m{\AA} or EW$_{6708}$\,=\,132\,m{\AA}), or less than 0.01\,dex at 6000\,K (models as previous), with the $\alpha$-enhanced opacities reducing the derived abundances relative to the non-enhanced models. This is comparable to the differences between some of our values and those of \cite{bm97}, although for other objects our abundances are smaller. We note that \cite{bm97} have taken the EWs published in several other surveys, averaged them where appropriate, and re-determined the Li abundances. They do not list which objects have been taken from which source, or how many observations have been averaged to produce each EW, so it is impossible to tell if any of the surveys used has provided systematically larger EWs than any other. They also report that EW measurements might be affected by non-Gaussian noise, such as scattered light, but do not mention what steps have been taken (if any) to minimise the impact of these effects, bearing in mind that they will also vary from sample to sample. Neglect of scattered-light correction would result in the lines being partially `filled in', reducing the observed EW and thus the derived abundances. 
\cite{rnb99} also commented on the non-uniformity of the sample, noting that the
inhomogeneity of the sample would have contributed to the spread observed in 
their results. We are confident that our homogeneous and consistent data 
reduction and analysis provide satisfactory results from 6708. 

\section{Discussion}
\label{sec-discuss}

\subsection{6104 {\em versus} 6708}
\label{sec-4v7}

Using consistent reduction and analysis techniques, we 
have measured 6104 in seven of our fourteen Pop.\,II stars and obtained upper 
limits in the others. We have measured 6708 in all the objects. 

\begin{figure}
\resizebox{9cm}{!}{\includegraphics{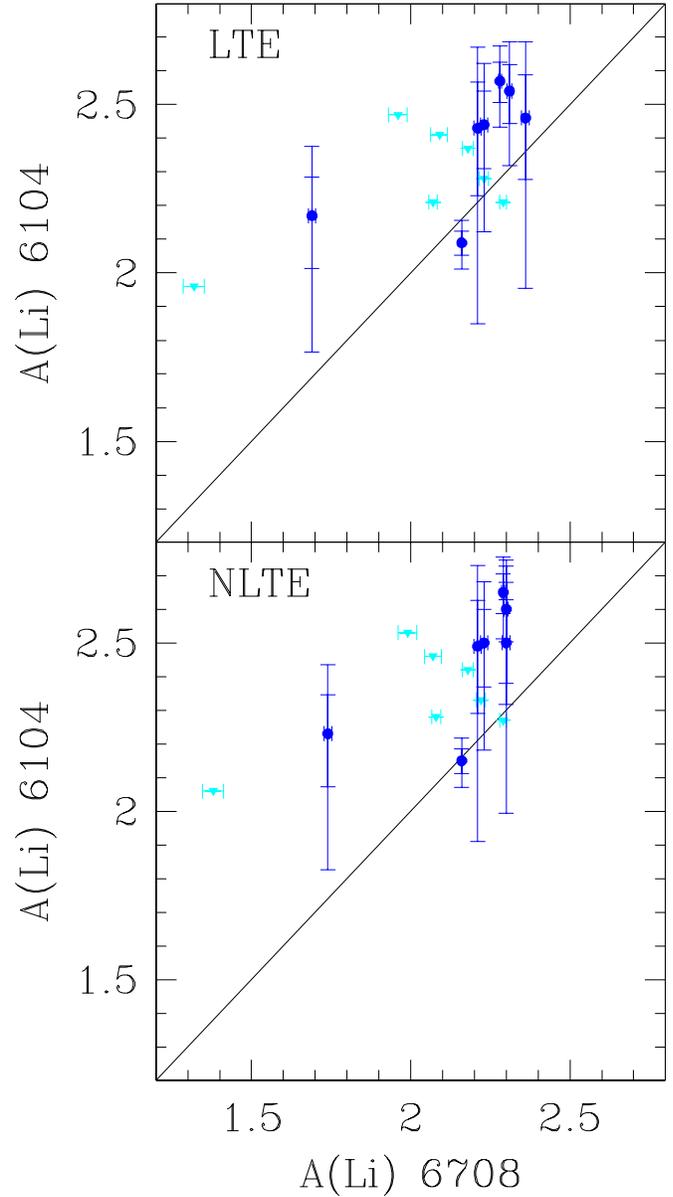}}
\caption{
A(Li)$_{6104}$ {\em versus} A(Li)$_{6708}$ for sample stars. 
The solid line is the 1-to-1 relation.
Since error bars which are linear in EW become non-linear in A(Li), especially for a line as
weak a 6104, we show both 1$\sigma$ and 2$\sigma$ error bars for the abscissa.
Upper limits are shown as inverted triangles.
{\it Upper panel}: LTE abundances.
{\it Lower panel}: NLTE abundances.
}
\label{fig-4v7}
\end{figure}

There is evidence of a discrepancy in Li abundance between the two lines 
for some of our sample objects. In Fig.~\ref{fig-4v7} which plots our 6104 
abundances {\em versus} those from 6708, it appears that the 
6104 abundances of some objects (e.g. HD\,219617 and G80-28) are higher than those from 6708. Figure 5 might also give the impression that we have uncovered a 
systematic discrepancy between the 6104 and 6708 abundances. However, part 
of this is due to the large number of upper limits, which make it 
impossible to draw such a conclusion. Indeed, as we describe below, there 
are some objects for which a large discrepancy can be ruled out. The discrepancy between 6104 and 6708 abundances is most apparent, reaching $\sim$0.5\,dex, at low values of A(Li) where, admittedly, there is more potential for overestimation of the 6104 equivalent width. At higher abundances the points lie closer to the one-to-one relation, but several are still $\sim$2$\sigma$ above it. 

A measurement of the 6104 line is more likely to be made not only in objects 
where the line {\em is} stronger, but also when noise acts to make it {\em appear} stronger. 
Because of this latter bias, it is not surprising that {\it some} of the 6104 
abundances exceed those of 6708. Nevertheless, it is disquieting that in 
three of the fourteen
cases, the abundance obtained from the 6104 line exceeds that from 6708 by more
than twice the formal error, {\em i.e.} $>$2$\sigma$. This can be seen in Fig.~\ref{fig-specfit4}: the 6708 abundance clearly {\em does not} fit the observations in some cases, most notably for HD\,219617 and G\,80-28. If the error distribution was 
Gaussian, and our error estimates were reliable, there would be only 
2\%\ probability of a 6104 abundance exceeding that from 6707 by 2$\sigma$. 
This would give rise to fewer than one discrepant case in fourteen. However, there are also cases where such discrepancies cannot exist: HD\,140283 (discussed below), HD\,108177 and HD\,94028 all have 6104 abundances which are consistent with the 6708 abundances. Overall, this leads to a very mixed picture where the abundances from the two lines agree for some objects, some objects have A(Li)$_{6104}$\,$>$\,A(Li)$_{6708}$, and some objects might have A(Li)$_{6708}$\,$>$\,A(Li)$_{6104}$(although the latter possibility is defined by only one object in our sample; the other upper limits neither contradict nor constrain this possibility). The lack of a clear picture of what is going on adds a further complication into any interpretation of the data, in that whatever causes the elevated 6104 abundances in some stars is not present in other objects.

\begin{figure*}
\resizebox{17cm}{!}{\includegraphics{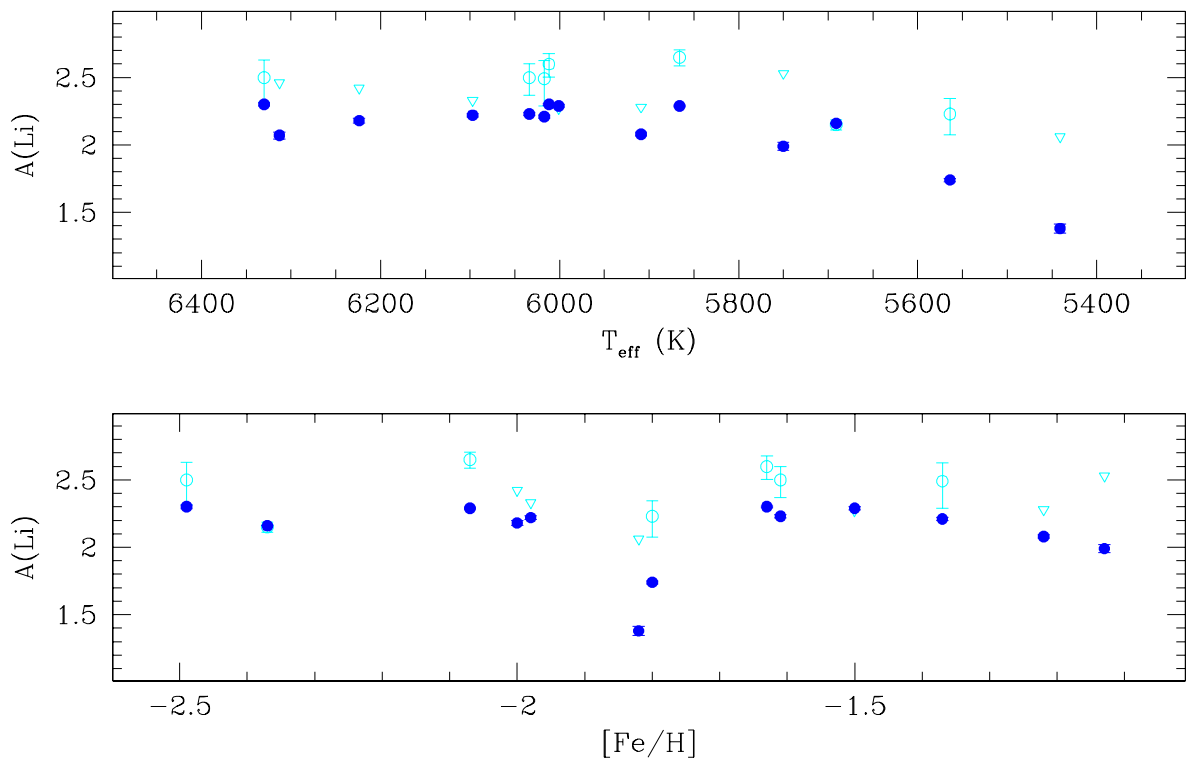}}
\caption{NTLE-corrected A(Li) {\em versus} $T_{\mathrm{eff}}$ (upper panel) and [Fe/H] (lower panel) for sample stars. Filled symbols represent 6708 abundances and open symbols show 6104 values. Inverted triangles indicate 6104 upper limits.}
\label{fig-teff-met}
\end{figure*}

We have plotted the abundances for the two lines against $T_{\mathrm{eff}}$ and 
[Fe/H] in Fig.~\ref{fig-teff-met}. In the case of the  A(Li)-{\em versus}-$T_{\mathrm{eff}}$ plot, the plateau can be seen for both lines, with similar abundances for objects with 
$T_{\mathrm{eff}} \geq$\,5800\,K. The 6104 abundances appear to show a separate plateau $\la$0.5\,dex above the 6708 plateau. As with Fig.~\ref{fig-4v7}, this impression is accentuated by the high number of upper limits, especially at low temperature, but concentrating only on the firm detections shows that the discrepancy is not an obvious function of effective temperature or metallicity. The Li abundances in the plateau objects all appear approximately constant with metallicity, although slopes in either direction could potentially be present. The behaviour of objects at the cool edge of the plateau is much as expected, with Li abundances from both lines decreasing with decreasing temperature, consistent with HD\,188510 and HD\,64090 being cool enough that some Li depletion is likely to have occurred during their lifetimes.

In order to investigate the abundance disparity further, more firm detections 
would be needed, although obtaining them is complicated by the increased 
difficulty in measuring such weak 6104 lines. Achieving higher $S/N$ exceeding
a few hundred not only requires longer and longer exposures, but also requires
highly repeatable flat-fields that correct for fringing and other defects to 
better than 0.1\%.

\subsection{HD 140283}
\label{sec-hd140}

The data for HD\,140283 do not show a discrepancy between the 6104 and 6708 abundances.
We also used our models to obtain abundances for HD\,140283 using the EWs and model 
parameters quoted in \cite{bm97} and \cite{bm98}. In contrast to the general
behaviour noted in Sect.~\ref{sec-4v7}, we find A(Li)$_{6708}$\,=\,2.16$\pm$0.01 
(from EW$_{6708}$\,=\,47.5$\pm$0.6\,m{\AA}), slightly higher than\,A(Li)$_{6104}$\,=\,2.10$^{+0.07}_{-0.08}$ (from EW$_{6104}$\,=\,1.8$\pm$0.3\,m{\AA})
but in any case in agreement within the errors. The abundance reported by 
\cite{bm98} for both lines in HD\,140283 is A(Li)\,=\,2.14$\pm$0.13, which is 
also in agreement with our calculated abundances.

\begin{figure}
\resizebox{!}{!}{\includegraphics{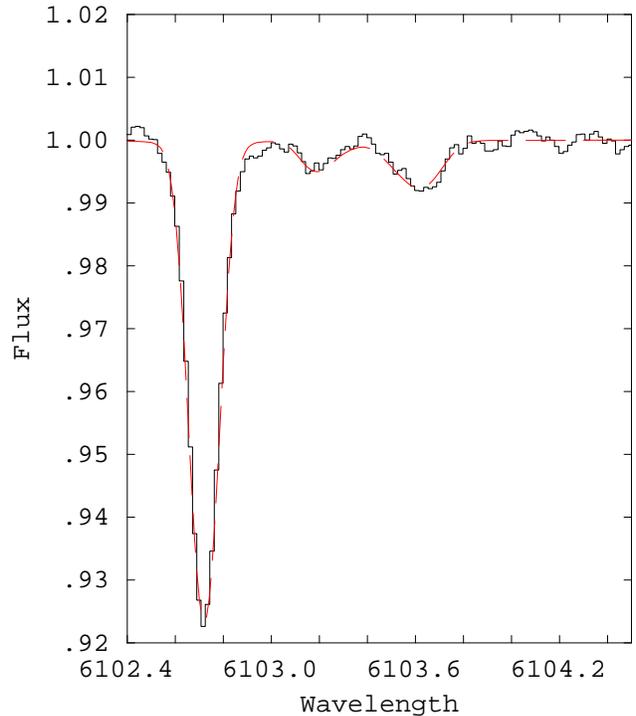}}
\caption{Observed (solid) spectrum and synthetic fit (dashed) to 6104 in HD\,140283.}
\label{fig-140zoom}
\end{figure}

Our observation HD\,140283 allows us to compare our results for this object directly with that obtained by BM98. The main differences between our analysis and theirs are: the significantly-higher $S/N$ of our spectrum ($\sim$1100 per 0.018{\AA} pixel, compared to BM98's 360 per 0.04{\AA} pixel); the $gf$ values used in the syntheses (see Table~\ref{table-linelist} for our values, and Sect.~2 of BM98); their use of $\alpha$-enhanced opacities {\em versus} our use of standard opacities (Sect.~\ref{sec-bmcomparison}); our use of ABO broadening parameters. We have also included an \ion{Fe}{ii} line in our synthesis of the 6104 region, although at the metallicity of the object this should make no difference. The analyses shared identical $T_{\mathrm{eff}}$, $\log g$, [Fe/H], $\xi$ and $v \sin i$ values. 
Despite the differences between the analyses, we have obtained abundances which are completely consistent with those reported by BM98, which is not too surprising since the differences are all in parameters which have only a small effect on abundance. Even so, it shows that there is a good degree of consistency between the two studies. Our fits to both 6708 and 6104 are remarkably close to the observations (Figs.~\ref{fig-specfit7} and \ref{fig-140zoom}), suggesting that our line profiles are reliable.

In concluding this subsection, we reemphasize that the 6104 lines are all very weak, and it is possible that we have underestimated the errors, but we discuss below the implications of a genuine discrepancy.

\subsection{Model atmospheres}
\label{sec-disc.atmospheres}

If the possible discrepancy between the 6104 and 6708 lines in some of the objects is real, what could explain it? Errors in $T_{\mathrm{eff}}$ of $\pm$100\,K lead to abundance 
{\it differences} between 6708 and 6104 of $\pm$0.03\,dex at 6000\,K, and 
$\pm$0.05\,dex at 5000\,K, with 6708 abundances changing more than 6104 
abundances. This has already been shown to be too small; a 1000\,K temperature 
increase would be required to bring the most discrepant lines into agreement 
(see Sect.~\ref{sec-abundanceerrors}), and the inclusion of other model 
uncertainties contributes less than 0.03\,dex. 

\begin{figure}
\resizebox{!}{!}{\includegraphics{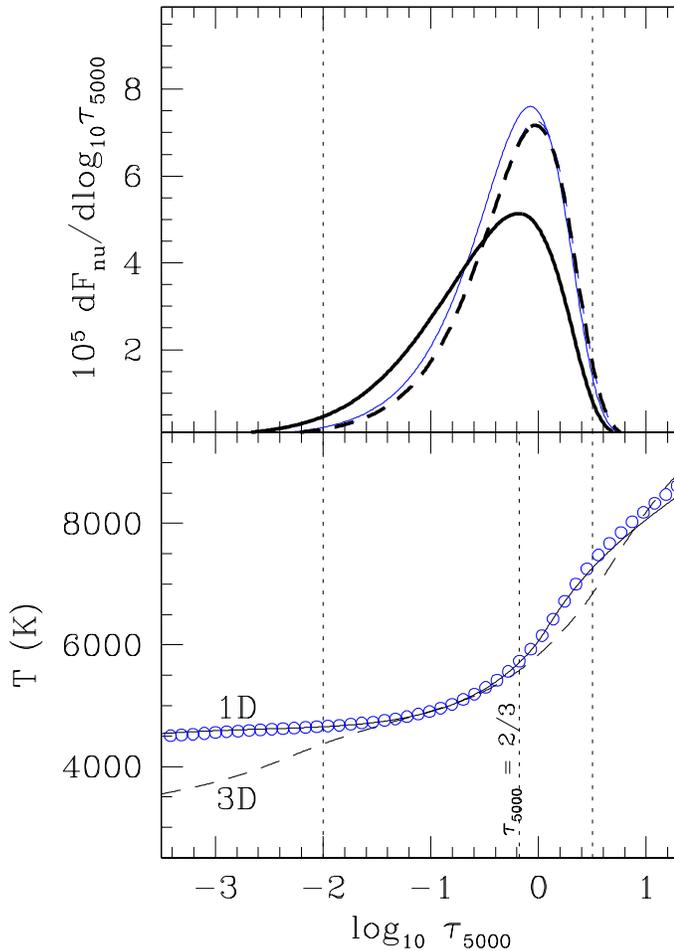}}
\caption{{\it Upper panel}: Flux contribution functions 
for our 1-D model of HD\,140283,
which shows the range of depths over which the spectrum forms. Shown are
contributions functions for 
the continuum near 6708\,\AA\ (thin solid curve),
the Li\,6708 line (heavy solid curve), 
the continuum near 6104\,\AA\ (thin dashed curve),
the Li\,6104 line (heavy dashed curve). 
(The contribution function plotted for each spectral line is a weighted average
of the contribution functions at several wavelengths within the line profile.)
{\it Lower panel}: Temperature profiles for our 1-D model (open circles),
and the 1-D (solid curve) and 3-D (dashed curve) models of 
Asplund et al. (1999). Dotted vertical lines are to guide the eye.
}
\label{fig-cf}
\end{figure}

An alternative explanation for our result is that the model atmospheres do not
adequately represent the stars we are studying, and a change in temperature
gradient within the stellar atmospheres might explain our observations. 
Asplund et al. (\cite{asplund99}) applied 3-D model atmospheres to two Pop.\,II 
stars, HD\,140283 and HD\,84937, noting the predicted effect on the Li abundances
relative to those from 1-D models. 3-D models alter the temperature 
stratification in a stellar atmosphere, with the greatest variations between 
1-D and 3-D models occurring at the inner and outer boundaries (see 
lower panel of Fig.~\ref{fig-cf}). The upper panel of Fig.~\ref{fig-cf} shows the flux contribution functions ${\mathrm{d}}(F_\nu)$/${\mathrm{d}} \log_{10}(\tau_{5000})$ computed for four wavelengths in the spectrum of HD\,140283, using our model atmosphere. These show the range of depths over which the spectrum forms at a given wavelength (e.g. Gray 1992, Chapter 13). The greater equivalent width of 6708 and its lower excitation potential both lead to it forming on average further out than 6104. 

The temperature differences between the 1-D and \nobreak{3-D} models of Asplund et al. are
fairly small at optical depths between $\frac{1}{30} < \tau_{5000} < \frac{2}{3}$. However, the 
lower temperatures of the 3-D models beyond this range, where part of the
flux at 6708 emerges, result in a stronger 
line forming in the 3-D models. (Switching to a 3-D model would also modify the 
contribution function from that shown). Asplund et al. found that, for a given
EW, the 6708-derived abundance in HD\,140283 would be 0.34\,dex lower if 
derived using 3-D LTE models rather than 1-D LTE atmospheres. 

Although 6104 could also be affected by the differences between 1-D and 3-D 
models, 6104 is so weak (EW\,=\,1.8\,m\AA) that it is barely distinguishable from 
the neighbouring continuum, and the magnitude of the effect is probably less for
this line. Asplund (priv. comm.) confirms this expectation. In this case, use of
3-D LTE models would decrease the 6708 abundance more than the 6104 abundance, 
{\em increasing} the abundance discrepancy identified in Sect.~\ref{sec-4v7} (if it
is real) between the two lines.

1-D NLTE effects, which have been corrected for in our results, have been discussed in three other studies of low-metallicity stars, namely Kurucz (\cite{kurucz95}), \cite{bm98}, and Molaro et al. (\cite{molaro95}). As can be seen from our Fig.~\ref{fig-4v7} the NLTE corrections are relatively small, affect both lines differently, and vary depending on temperature, metallicity and Li abundance. In all cases, however, application of the NLTE correction acts to {\em increase} the 6708 and 6104 abundance difference, rather than reducing it.

Asplund et al's (1999) suspicion that overionisation of Li in 3-D models might 
reverse the 0.3\,dex difference initially found for the 6708 line between the 
1-D LTE and 3-D LTE models proved to be correct (Asplund 2000). However, 
corrections for overionisation may be expected to affect both the 6104 and 6708
lines, though not necessarily equally (due to the different depths of 
formation). It is clear that detailed calculations using 3-D NLTE atmospheres 
will be required for each programme star to fully evaluate the competing effects
and their different action on the two lines.

To produce the discrepancy via temperature gradients alone, what would be 
required is a process which changes the temperature gradient in opposite ways 
for the two lines, such that it increases the temperature in the outer region 
where 6708 is formed, thus weakening that line and leading to an underestimate
of its abundance, and decreasing the temperature in the deeper region where 6104
forms. As noted above, this is opposite to the behaviour seen in the 3-D 
atmospheres of Asplund et al. (\cite{asplund99}). 

\subsection{Binarity}
\label{sec-binarity}

There are no clear systematic effects of binarity on Li abundances derived from either line. Of our seven detections, three are in objects classified as binaries by Carney (\cite{carney83}): HD\,84937, HD\,219617 and HD\,188510; both HD\,84937 and HD\,188510 have also been classified as definite binaries by Stryker et al. (\cite{stryker85}). One object in our sample, BD+21\,607, is near the significance criterion for binarity (Stryker et al. \cite{stryker85}), and the remaining three have not been classified as binaries by either study. 
The upper limits do nothing to clarify the picture with respect to binarity either: LP\,608-62, HD\,94028, HD\,108177 and HD\,210891 have all been classified as either definite or probable binaries by Carney or Stryker et al., while the remainder are probably not binaries.
Clearly some other solution is required to explain the abundance discrepancy between the two lines. 

\subsection{Contrast with Pop. I stars}
\label{sec-popi}

In a related paper (Ford et al. \cite{ford02}) we found evidence that 6104 gave higher Li abundances than 6708 for some young Pleiades G/K dwarfs, whilst for others there seems to be reasonable agreement. This is very similar to the situation we find here for Pop. II stars.

For the Pleiades stars we were able to bring the abundances estimated from the two lines into agreement by introducing cool starspots in to the atmospheric models. We did this in a simple way, by modelling the atmosphere as two one-dimensional components with differing temperatures and surface areas. In young Pop. I stars there is plenty of evidence (from Doppler imaging and light-curve modulation) that such spots exist, and so it is sensible to incorporate them in the models.  Our conclusion was that plausible spot coverages and temperatures could explain the abundance discrepancies we saw in one-component models, although we had insufficient data to determine whether the stars actually did possess spots with the right properties.

We do not expect to find large, magnetically-generated spot regions associated with Pop. II stars. They are old, have spun down and their convection zones are much thinner than the Pop. I objects we considered in the Pleiades. For these reasons we do not think it appropriate to consider cool starspots as an explanation for the abundance discrepancies we see in Pop. II stars. That we still see a discrepancy (in some but not all stars) in Pop. II objects is likely to be an indication that something else is still wrong with our atmospheric models. It may also be the case that these additional problems are also present in Pop. I stars, and that the starspot interpretation is only part of the story. Indeed, even after the introduction of starspots on to the  Pop. I stars we were unable to reduce the large dispersion that is seen in their Li abundances.

\subsection{Effects on cosmological parameters: $\eta$ and $\Omega_{\mathrm{B}}$}
\label{sec-cosmology}

If we consider that the 6104-derived abundances are less likely to be affected 
by multi-dimensional model-atmosphere variations than those from 6708, then the
plateau abundance inferred from our 6104 {\it detections} would be A(Li)$\sim$2.5. (We do not include possible depletion factors in the values quoted below, although our 6104 abundance is the same as that derived by Salaris \& Weiss (\cite{salaris01}) from their diffusion models, constrained by observations of 6708). 
This would lead to two possible values for the baryon-to-photon ratio ($\eta$: values given here are $\eta_{10}\,=\,10^{10}\eta$) of either $\sim$1.0 or $\sim$7.0, leading to values of 3.7$\times\,10^{-3}h_{\mathrm{o}}^{-2}$ or 25.5$\times\,10^{-3}h_{\mathrm{o}}^{-2}$ for $\Omega_{\mathrm{B}}$, the Universal baryon density (where $h_{\mathrm{o}}$\,=\,$\frac{H_{\mathrm{o}}}{100}$\,km\,s$^{-1}$\,Mpc$^{-1}$ and the microwave background temperature is 2.73\,K). Values for A(Li)$_{\mathrm{plateau}}$\,=\,2.09 (Ryan et al. \cite{ryan00}) would be: $\eta_{10}$\,$\sim$\,1.2 or 6.0; $\Omega_{\mathrm{B}}$\,$\sim$\,4.4$\times10^{-3}h_{\mathrm{o}}$ or 21.9$\times10^{-3}h_{\mathrm{o}}^{-2}$, with the $\sim$0.4\,dex change in A(Li) affecting $\Omega_{\mathrm{B}}$ by $\sim$16\%. In either case, this assumes that the plateau abundance is indicative of the primordial Li abundance. However, this assumption may be unsound. The work of \cite{rnb99} found that the plateau slopes with metallicity, suggesting that the material from which the stars formed was gradually being metal enhanced over the time when Pop. II objects were being created. Despite the difficulties in measurement of weak lines, abundances from low-metallicity stars, which we assume to have formed earliest, still provide the best starting point for determinations of Li$_{\mathrm{p}}$ and related cosmological parameters.

\section{Summary}
\label{sec-summary}

We have obtained data for a sample of 14 Pop. II stars with high signal-to-noise ratios, which were all consistently reduced and analysed.
\begin{itemize}
\item We measured the 6708{\AA} \ion{Li}{i} resonance line in all our sample objects. We compared our EWs and abundances with those of Bonifacio \& Molaro (\cite{bm97}), finding good agreement between the values at a 99\% confidence level.
\item We also measured the 6104{\AA} \ion{Li}{i} subordinate line in seven of the stars, obtaining upper limits for the rest of the sample;
\item Abundances from the 6104{\AA} line hinted at a systematically higher Li abundance than those from the 6708{\AA} line, with some stars discrepant by up to $\sim$0.5\,dex, while in others the discrepancy is no larger than $\sim$0.1\,dex. This trend was very weak, and a large preponderance of upper limits prevent any firm conclusions from being drawn; 
\item This difference cannot be explained by including NLTE-corrections, which actually make the discrepancy larger. Binarity does not appear likely to affect the abundance difference;	
\item The 3-D model results of Asplund et al. (\cite{asplund99}) for Pop. II stars HD\,84937 and HD\,140283 suggest that the use of multi-dimensional atmosphere models would most likely increase any abundance discrepancy between the lines relative to that inferred from one-dimensional models. Our results can be used to place strong constraints on the 3-D modelling of these stars, perhaps indicating areas where further improvements could be made (e.g. improved NTLE-modelling, inclusion of magnetic fields). As they stand, however, we do not believe that 3-D models can explain our results.
\end{itemize}
\begin{acknowledgements}
The William Herschel and Isaac Newton Telescopes are operated on the island of La Palma by the Isaac Newton Group in the Spanish Observatorio del Roque de los Muchachos of the Instituto de Astrofisica de Canarias. The AAO operates the Anglo-Australian and UK Schmidt telescopes on behalf of the astronomical communities of Australia and the UK. To this end the Observatory is funded equally by the Australian and British Governments. Subaru is an 8.2 meter optical-infrared telescope at the summit of Mauna Kea, Hawaii, operated by the National Astronomical Observatory of Japan (NAOJ) with the support of the Ministry of Education, Culture, Sports, Science, and Technology. We extend our gratitude to Martin Asplund for his helpful advice and comments, and for kindly providing data for Fig.~\ref{fig-cf}. 
SGR was supported by grant PPA/0/S/1998/00658 from the UK Particle Physics and Astronomy Research Council (PPARC). DJJ thanks the PPARC for a post-doctoral research fellowship, and the Royal Society for a European research grant. DJJ would like to acknowledge the continued positive influences of Mrs J. Pryer. 
The authors acknowledge the travel and subsistence support of PPARC. AF and JRB were funded by PPARC postgraduate studentships. Computational work was performed on the Keele, Open University and St Andrews nodes of the PPARC-funded Starlink network. This research has made extensive use of NASA's Astrophysics Data System Abstract Service.
\end{acknowledgements}

\end{document}